\documentclass{jpp} % jpp
\usepackage{graphicx}
\usepackage{epstopdf, epsfig}
\usepackage{hyperref}
\usepackage{mathrsfs}
\usepackage[normalem]{ulem}
\usepackage{xcolor}
\newcommand{\fv}[1]{{\color{red}#1}}%FV
\shorttitle{Kinetic Alfv\'en waves and velocity shear}
\shortauthor{T. Maiorano et al.}

\title{Kinetic Alfv\'en wave generation by velocity shear in collisionless plasmas}

\author{T. Maiorano\aff{1}, A. Settino\aff{1}, F. Malara\aff{1}
 \corresp{\email{francesco.malara@fis.unical.it}}, 
       O. Pezzi\aff{2,3}, F. Pucci\aff{4}, \and F. Valentini\aff{1}}
\affiliation{
\aff{1}Dipartimento di Fisica, Universit\`a della Calabria, 87036 Rende (CS), Italy
\aff{2}Gran Sasso Science Institute, Viale F. Crispi 7, I-67100 L. Aquila, Italy
\aff{3}INFN/Laboratori Nazionali del Gran Sasso, Via G. Acitelli 22, I-67100 Assergi (AQ), Italy
\aff{4}Centre for Mathematical Plasma Astrophysics, Department of Mathematics, KU Leuven, Celestijnenlaan 200B, 3001 Leuven, Belgium}

\begin{document}

\maketitle

\begin{abstract}
%{\tm{Teresa}}, {\as{Adriana}}, {\fm{CiccioM.}}, {\fv{Franco}}, {\op{Oreste}}, {\fp {CiccioP.}}.
%
%A long-standing outcome in space plasmas observations is the evidence of alfvenic-like fluctuations propagating in the solar wind. 
%Recently, the Parker Solar Probe mission pointed out a new remark of these waves at a very short heliospheric distance.
%The nonlinear interaction of such perturbations leads to the coupling of different scales and an energy cascade towards smaller and smaller wavelengths is observed. 
%A crucial role in such turbulence dynamics is played by the interaction of the fluctuations with large scale inhomogeneities. 
%
The evolution of a linearly-polarized, long-wavelength Alfv\'en wave --propagating in a collisionless magnetized plasma with a sheared parallel-directed velocity flow-- is here studied by means of two-dimensional hybrid Vlasov-Maxwell (HVM) simulations. The unperturbed sheared flow has been represented by an exact solution of the HVM set of equations (Malara {\it et al.}, Phys. Rev. E 97, 053212), this avoiding spurious oscillations that would arise from the non-stationary initial state and inevitably affect the dynamics of the system. We have considered the evolution of both a small and a moderate amplitude Alfv\'en wave, in order to separate linear wave-shear flow couplings from kinetic effects, the latter being more relevant for larger wave amplitudes. The phase-mixing generated by the shear flow modifies the initial perturbation, leading to the formation of small-scale transverse fluctuations at scales comparable with the proton inertial length. By analyzing both the polarization and group velocity of perturbations in the shear regions, we identify them as Kinetic Alfv\'en Waves (KAWs). In the moderate amplitude run, kinetic effects distort the proton distribution function in the shear region. This leads to the formation of a proton beam, at the Alfv\'en speed and parallel to the magnetic field. Such a feature, due to the parallel electric field associated with KAWs, positively compares with solar-wind observations of suprathermal ions' populations, suggesting that it may be related to the presence of ion-scales KAW-like fluctuations.
%\as{non sono sicura di quest'ultima frase. In pratica l'ho presa dalla conclusione, mi sembrava carina per concludere l'abstract}

%This file contains instructions for authors planning to submit a paper to the {\it Journal of Plasma Physics}. These instructions were generated in {\LaTeX} using the JPP style, so the {\LaTeX} source file can be used as a template for submissions. The present paragraph appears in the \verb}abstract} environment. All papers should feature a single-paragraph abstract of no more than 250 words, which provides a summary of the main aims and results.
\end{abstract}

\section{Introduction}
In several natural contexts low-frequency fluctuations display properties which are typical of an Alfv\'enic state: velocity and magnetic field perturbations are highly correlated (positively or negatively), while density and magnetic field intensity are affected by much less intense fluctuations. 
This situation is typically found in solar wind turbulence, mainly in low-latitude fast-speed streams, or in the high-latitude wind \citep[e.g.,][]{belcher71,bruno13}. Recently, measures performed by the Parker Solar Probe spacecraft at small heliocentric distances have shown fluctuations in the solar wind emanating from an equatorial coronal hole with strong Alfv\'enic correlation, even in cases of very large amplitudes (the so-called switchbacks)\citep{bale19}.
Another example is that of velocity fluctuations propagating along the magnetic field at the Alfv\'en speed detected in the solar corona \citep{tomczyk07,tomczyk09}, which have been interpreted as Alfv\'en waves.

In the framework of the magnetohydrodynamic (MHD) turbulence, such perturbations {\fv interact nonlinearly} producing a cascade which moves fluctuating energy at increasingly smaller scales. In the presence of a background magnetic field ${\bf B}_0$, nonlinear interactions preferentially take place in the directions transverse to ${\bf B}_0$ \citep{shebalin83,carbone90,oughton94}. Therefore, an anisotropy develops in the spectral space, with perpendicular wavevectors dominating over parallel ones at small scales. Indeed, solar-wind observations have shown the presence of a significant population of quasi-perpendicular wavevectors \citep{matthaeus86,matthaeus90,carbone95,milano01,dasso05,matthaeus12,oughton15}. A further element, which can play a role in the turbulent dynamics, is given by the interaction of fluctuations with inhomogeneities associated with large scale structures, such as pressure-balanced structures or velocity shears \citep{ghosh98}. For instance, this happens in the solar wind around the heliospheric current sheet \citep{malara96a} or at the interface between fast and slow speed streams \citep{roberts91,roberts92}. Finally, fluctuations in the turbulent spectrum at scales of the order or lower than ion scales (ion inertial length and/or ion Larmor radius) are affected by dispersive and kinetic phenomena. In a collisionless plasma, these kinetic processes generate out-of-equilibrium features in the particle distribution function: ion temperature anisotropy has been routinely observed in the solar wind \citep[e.g.,][]{hellinger06}; the proton distribution function in the solar wind and in the Earth's magnetosphere can include a beam directed in the direction parallel to the local magnetic field \citep{goodrich76,marsch82,marsch06,sorriso19}, with a drift speed of the order of the local Alfv\'en velocity \citep{goldstein00,tu04}. Physical processes which can lead to the formation of this beam have also been considered \citep{araneda08,valentini08,matteini10,valentini11a,valentini11b,nariyuki14a}. More recently, a further level of velocity-space complexity, namely the presence of an enstrophy cascade towards smaller velocity-space scales, has been pointed out either in magnetosheath' observations \citep{servidio17} and in numerical simulations \citep{pezzi18,cerri18}, which are also useful to investigate the nature of cross-scale correlations between the inertial-range turbulent energy cascade and the small-scale kinetic processes in collisionless plasmas \citep{sorriso18}.

A full description which takes into account the above features, namely, Alfv\'enic correlations, spectral anisotropy, the role of background inhomogeneities, and kinetic effects at ion scales, is a complex task. However, some insight can be gained by means of simplified approaches. In particular, in this paper we will focus on a specific aspect: the interaction between an Alfv\'enic perturbation and a bulk velocity inhomogeneity transverse to the background magnetic field, which produces small scales of the order of the ion inertial length in the transverse direction. This effect somehow mimics the anisotropic small-scale generation taking place in a turbulence, where large scales evolve on characteristic times which are larger than that of smaller scales.

Such a problem has been already analyzed in detail within the MHD description, where two distinct mechanisms are at work: {\it phase-mixing}, in which differences in phase speed produce a progressive de-phasing that bends wavefronts and increases the transverse component of the wavevector; and {\it resonant absorption}, in which the wave energy progressively concentrates in thin layers where a resonance condition is satisfied. These processes have been studied both by normal mode approaches \citep{kappraff77,mok85,steinolfson85,davila87,hollweg87,califano90,califano92} and by considering the evolution of an initial disturbance \citep{lee86,malara92,malara96b}. Localized pulses \citep{kaghashvili99,tsiklauri02,tsiklauri03} have also been considered. The propagation of MHD waves in inhomogeneous magnetic fields containing null points has also been studied \citep{landi05,mclaughlin11,pucci14},
finding a fast formation of small scales perpendicular to the ambient magnetic field. In 3D inhomogeneous
equilibria this process has been considered in the small wavelength limit \citep{similon89,petkaki98,malara00}, also within the problem of coronal heating \citep{malara03,malara05,malara07}.

When the wavelength of the perturbation has decreased until reaching values comparable with the ion inertial length, both dispersive and kinetic effects, which are neglected in the MHD approach, become important. In particular, an Alfv\'en wave with a wavevector which is (i) quasi-perpendicular to the background magnetic field, and (ii) of the order of the inverse of the ion inertial length, is generally indicated as ``kinetic Alfv\'en wave" (KAW). KAWs have received considerable attention because observations have shown that the polarization of fluctuations at kinetic scale in the solar wind is consistent with the presence of KAW-like fluctuations \citep{chen13}. Since the MHD cascade favors nearly perpendicular wavevectors, the expectation within a wave perspective would be that fluctuations having a character resembling KAWs were naturally present at scales of the order of the ion inertial length. An extensive analysis of the KAW physics can be found in \citet{hollweg99} (see also references therein for a more complete view on the subject). Many solar wind observational analyses \citep{bale05,sahraoui09,podesta12,salem12,chen13,kiyani13}, theoretical works \citep{howes08a,schekochihin09,sahraoui12} as well as numerical simulations \citep{gary04,howes08b,tenbarge12} have suggested that fluctuations near the end of the MHD inertial cascade range may consist primarily of KAWs and that such fluctuations can
play an important role in the dissipation of turbulent energy.
Formation of KAW-like small-amplitude fluctuations has also been observed in numerical simulations of collisions between two counter-propagating Alfv\'enic wavepackets \citep{pezzi17a,pezzi17b,pezzi17c}, a process which mimics nonlinear interactions among localized eddies in turbulence. Moreover, Landau damping and wave-particle resonant interactions can take place in KAWs \citep{vasconez14}. Due to a nonvanishing electric field parallel component associated with KAWs, these waves have also been considered in the problem of particle acceleration \citep{voitenko04,decamp06}.
Particle acceleration in {\it phase-mixing} of Alfv\'en waves in a dispersive regime has been studied both in 2D \citep{tsiklauri05,tsiklauri11} and in 3D \citep{tsiklauri12} configurations. Finally, instabilities generating KAWs in a plasma with transverse density modulations have been considered by \citet{wu13}. Similar ideas involving dissipative mechanisms related to interaction of Alfv\'en waves or KAWs and {\it phase-mixing} have been examined in the context of the magnetospheric plasma sheet \citep{lysak11} and in coronal loops \citep{ofman02}. It has been shown that ion-scale shear Alfv\'en waves can be excited by ion beams in the solar wind \citep{hellinger11,hellinger13}, and these can contribute to the formation of KAWs \citep{nariyuki14b}. The possible role played by KAW-like fluctuations in heating electrons in coronal turbulence has been considered \citep{malara19}.

Recently, the {\it phase-mixing} of Alfv\'enic perturbations propagating in a pressure-balanced magnetic field transverse inhomogeneity has been numerically studied, by comparing the results of Hall-MHD and kinetic simulations \citep{vasconez15,pucci16,valentini17}. In these simulations the perturbation wavelength decreases, due to the interaction with the inhomogeneous background, until it reaches values of the order of the ion inertial lengh; the properties of the initial perturbation gradually change, eventually leading to an efficient generation of KAWs inside the inhomogeneity region. Moreover, it has been observed that such KAWs sensibly modify the initially Maxwellian ion distribution function \citep{vasconez15,valentini17}, producing temperature anisotropy, as well as localized ion beams which move along the magnetic field with a speed close to the local Alfv\'en velocity. 

In the present paper we study a similar problem, namely, the evolution of a large-scale Alfv\'enic perturbation propagating in a collisionless plasma with a sheared velocity field and a uniform magnetic field parallel to the velocity. Shearing flows in plasmas with a quasi-parallel magnetic field can be found, for instance, in the interaction region between fast and slow streams of the solar wind \citep{bruno13}, or in astrophysical jets \citep{hamlin13}. We consider an unperturbed configuration where the width of the velocity shear is of the order of few ion inertial lengths and we use a hybrid Vlasov-Maxwell approach to describe the system evolution. Our numerical simulations show that, also in the present case, {\it phase-mixing} acting on the Alfv\'enic perturbation increases the perpendicular wavenumber, until KAWs develop inside the velocity shear regions. For sufficiently large amplitudes, such waves modify the ion distribution function, locally generating a particle beam propagating along the magnetic field at the Alfv\'en velocity.

The plan of the paper is the following: in Section 2 we present the equations describing our model, including the form of the stationary solution and of the perturbation; in Section 3 we describe the results of the simulations; and in Section 4 we give the conclusions.

\section{The model}
We consider a collisionless, fully-ionized, magnetized plasma composed by protons and electrons.
%\fv{uniformiamo e parliamo solo di protons allora, dappertutto} \fm{non sono d'accordo: le considerazioni fatte nell'Introduzione sono generali e non necessariamente valgono nel caso in cui gli ioni sono protoni. Invece, da qui in poi inizia il modello usato in questo articolo, fatto di protoni ed elettroni} 
We want to describe phenomena taking place at spatial scales larger than or of the order of the proton inertial length $d_p=c_A/\Omega_{p}$ and/or of the proton Larmor radius $\rho_p=v_{th,p}/\Omega_{p}$. In previous definitions, $c_A=B/(4\pi m_p n)^{1/2}$ and $v_{th,p}=(\kappa_B T_p/m_p)^{1/2}$ are respectively the Alfv\'en and the proton thermal speeds, related through the proton $\beta_p=2 v_{th,p}^2/c_A^2$, while $\Omega_{p}=eB/m_p c$ is the proton cyclotron frequency. Moreover, $B$ is the magnetic field; $n$ is the density (assumed to be equal for protons and electrons); $T_p$, $m_p$ and $e$ are the proton temperature, mass and charge, respectively; $\kappa_B$ is the Boltzmann constant and $c$ is the speed of light.
%\op{[Definerei qui il proton beta]}\fm{Franco, ci puoi pensare tu? Io col beta dei protoni mi scasino sempre...}

\subsection{Equations of the model}
We employ the Hybrid Vlasov-Maxwell (HVM) model, where protons are kinetically described by the Vlasov equation, while electrons are treated as a massless fluid. As usual in the HVM approach, we assume that the electron fluid is isothermal: $T_e={\rm const}$. The equations describing the HVM model, in dimensionless units, are the following:
\begin{eqnarray}
\frac{\partial f}{\partial t} + {\bf v} \cdot \nabla f + \left( {\bf E} + {\bf v}\times {\bf B} \right) \cdot \frac{\partial f}{\partial {\bf v}} =0 \label{vlasov}\\
{\bf E} = -{\bf u} \times {\bf B} + \frac{1}{n} \left( {\bf j}\times {\bf B}\right) - \frac{1}{n} \nabla p_e \label{ohm}\\
\frac{\partial {\bf B}}{\partial t} = -\nabla \times {\bf E}; \;\;\;\; {\bf j}=\nabla \times {\bf B}; \;\;\;\; p_e = n T_e \label{three_eq}
\end{eqnarray}
where, $f=f({\bf x}, {\bf v},t)$ is the proton distribution function (DF), $n=n({\bf x},t)$ and ${\bf u}={\bf u}({\bf x},t)$ are the density and proton bulk velocity, respectively:
\begin{equation}\label{n_u}
n({\bf x},t) = \int f({\bf x}, {\bf v},t) d^3 {\bf v}; \;\;\;\;
{\bf u}({\bf x},t) = \frac{1}{n({\bf x},t)} \int {\bf v} f({\bf x}, {\bf v},t) d^3 {\bf v}
\end{equation}
In the above equations the magnetic field ${\bf B}$ is normalized to a typical value ${\tilde B}$; the density $n$ is normalized to a typical value ${\tilde n}$;  velocities ${\bf v}$ and ${\bf u}$ are normalized to the typical Alfv\'en speed ${\tilde c}_A={\tilde B}(4\pi m_p {\tilde n})^{-1/2}$; the electric field ${\bf E}$ is normalized to ${\tilde E}=({\tilde c}_A/c){\tilde B}$; the time $t$ is normalized to the typical proton gyration time ${\tilde \Omega}_p^{-1}$, with ${\tilde \Omega}_p=e{\tilde B}/(m_p c)$; space variables are normalized to the typical proton inertial length ${\tilde d}_p = {\tilde c}_A/{\tilde \Omega}_p$; the current density ${\bf j}$ is normalized to the value ${\tilde j}=c{\tilde B}/(4\pi {\tilde d}_p)$; and the electron temperature $T_e$ is normalized to the typical temperature ${\tilde T} = {\tilde B}^2/(4\pi \kappa_B {\tilde n})$. 
%\as{scusate ma non mi tornano molto le normalizzazioni. E e B non sono normalizzate alla stessa quantit\'a perch\'e ${\tilde B}=m_pc \Omega_{cp}/e$ mentre ${\tilde E}=m_p c_A \Omega_{cp}/e={\tilde B}c_A/c$ almeno una velocit\'a di Alfven di mezzo. Inoltre non dovrebbe essere contrario il ragionamento? Vale a dire, si fissano le quantit\'a che normalizzano lunghezza, densit\'a e tempo e poi da queste derivano le normalizzazioni di v, E, B e T. O sbaglio? }
In what follows, all the results will be expressed in terms of the above-defined dimensionless variables.

The system of equations (\ref{vlasov})-(\ref{n_u}) is solved by means of the HVM numerical algorithm \citep{valentini07}. The spatial domain is 2D and is defined by $D_{\bf x}=\left\{ (x,y)\right\} = \left[ 0,L\right] \times \left[ 0,L \right]$, $L=16\pi$ being the domain size, while the 3D domain in the velocity space is defined by $D_{\bf v} = \left\{ (v_x,v_y,v_z), -7v_{th,p} \le v_i \le 7v_{th,p}\; , i=x,y,z\right\}$, where $v_{th,p}$ is a typical dimensionless proton thermal speed, defined in the next section. 
%\fm{Ma finora $v_{th,p}$ non \`e stata ancora definita: che ne dite di mettere la frase seguente: where $v_{th,p}$ is the proton thermal speed associated with the stationary state, far from the shear layers (see below)}. 
Periodic boundary conditions are imposed on the boundaries of the spatial domain $D_{\bf x}$ for all quantities, while the DF is imposed to vanish at the boundaries of the velocity space domain $D_{\bf v}$. More details on the numerical method can be found in \citet{valentini07}.
%Concerning the initial condition, at the initial time $t=0$ each quantity $f$ is decomposed as $f_0 + f_1$, where $f_0$ corresponds to the stationary configuration part, and $f_1$ to the perturbation. {\fp{[suggerisco di cambiare il nome della quantit\`a generica perch\'e ricorda quello della funzione di distribuzione e potrebbe confondere il lettore.}} \op{[concordo]}\fv{Eliminerei del tutto questa frase, \'e inessenziale}

\subsection{Stationary configuration}
The stationary configuration corresponds to a magnetized plasma with a shearing flow, where the magnetic field ${\bf B}_0$ is uniform and directed parallel to the sheared bulk velocity ${\bf u}_0$. Building such kind of configuration is straightforward in the case of MHD, but it is more complex within a kinetic approach. Explicit solutions have been found by \citet{roytershteyn08} in the fully kinetic case, and by \citet{malara18} within the HVM approach. Here, we will use the solution of the HVM case, which is briefly described in the following; more details can be found in \citet{malara18}. 

We consider a Cartesian reference frame, where the uniform magnetic field is directed along the $y$ axis: ${\bf B}_0=B_0 {\bf e}_y$ ($B_0=1$ in code units), while the bulk velocity is directed along $y$ as well, but varies in the $x$ direction: ${\bf u}_0=u_0(x) {\bf e}_y$, ${\bf e}_x$ and ${\bf e}_y$ being the unit vectors in the $x$ and $y$ directions, respectively. The electric field is vanishing: ${\bf E}_0=0$, therefore protons move along helical trajectories, with the helix axis parallel to the $y$ direction. In this configuration, the motion invariants are: the particle kinetic energy $\mathscr{E}$, the parallel velocity component $v_y$, and the $x$-position of the particle guiding center  $x_c=x-v_z/\Omega_{p}$, ($\Omega_{p}=1$ in code units). We build a proton DF $f_0$ in terms of the above constants of motion, which is similar to a Maxwellian, shifted in the $v_y$ direction by a quantity $U(x_c)$, where $U(\cdot )$ is an arbitrary function which contributes to determine the profile $u_0(x)$ of the bulk velocity. The explicit form of the DF is given by the following combination of the motion constants:
\begin{equation}\label{f0}
f_0(x,v_x,v_y.v_z)=\frac{n_0}{(2\pi )^{3/2} v_{th0,p}^3} \exp \left[ -\frac{1}{2 v_{th0,p}^2} 
\left\{ v_x^2 + \left[ v_y - U\left( x - \frac{v_z}{\Omega_{p}}\right) \right]^2 + v_z^2 \right\} \right]
\end{equation}
where the meaning of the constants $n_0$ and $v_{th0,p}$ is specified below.
%where $n_0$ and $v_{th,p}$ are other constants whose meaning is specified below \fv{$v_{th,p}$ l'abbiamo gi\'a definita sopra, \'e la velocit\'a termica, qui stiamo dando l'idea che sia qualcosa di diverso e questo crea confusione} \fm{In realt\`a sopra non era stata definita...Vedi il mio commento a riguardo (sopra)}. 
Since $f_0$ is expressed only in terms of single-particle motion constants, it is an exact stationary solution of the Vlasov equation (\ref{vlasov}), provided that both ${\bf B}$ and ${\bf E}$ remain stationary. Calculating the moments of the DF, we find that the associated density is uniform: $n\equiv \int f_0\, d^3{\bf v}=n_0={\rm const}$, and the bulk velocity is directed along $y$: $u_{0x}=u_{0z}=0$. 
As a consequence, since $T_e={\rm const}$, we have $\nabla p_e=0$ (third equation (\ref{three_eq})). Moreover, being ${\bf u}_0$ parallel to ${\bf B}$, and ${\bf B}$ uniform, the generalized Ohm's law (\ref{ohm}) implies that ${\bf E}=0$. In turn, this gives a vanishing time derivative of ${\bf B}$ (Faraday's law (\ref{three_eq})). In summary, both the electric and magnetic fields are stationary and the considered configuration represents a stationary solution for the entire set of HVM equations.

In the particular case where the function $U(\cdot)$ is constant, the expression (\ref{f0}) reduces to a shifted Maxwellian; therefore, the constant $v_{th0,p}$ can be interpreted as the thermal velocity of protons in regions far from the velocity shear, where the distribution function (\ref{f0}) approaches a shifted Maxwellian.
The only nonvanishing bulk velocity component $u_{0y}$ depends on the function $U(\cdot)$ through the expression \citep{malara18}:
\begin{equation}\label{u0y}
u_{0y}(x)=\frac{1}{(2\pi)^{1/2} v_{th0,p}} \int_{-\infty}^{\infty} U\left( x-\frac{v}{\Omega_{p}}\right) 
\exp \left( -\frac{v^2}{2 v_{th0,p}^2} \right) dv
\end{equation}

Therefore, the bulk velocity profile $u_{0y}(x)$ does not coincide with $U(x)$, except in the case $U(x)=U_0={\rm const.}$, when $u_{0y}(x)=U_0$. In general, denoting with $\Delta$ the scale of variation of the function $U(\cdot)$, it can be shown that $u_{0y}(x)\simeq U(x)$, if $\Delta \gg R_p$, where $R_p$ is the dimensionless proton Larmor radius. Considering the opposite limit $\Delta < R_p$, it can also be shown that, in the considered configuration, the actual width of a velocity shear cannot be smaller than the proton Larmor radius \citep{malara18}.

\begin{figure}
%  \centerline{ 1{figure1.eps}}
  \centerline{\includegraphics[width=7cm]{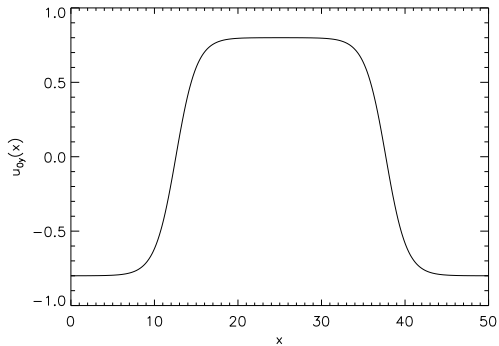}}
  \caption{Profile of the stationary-state bulk velocity $u_{0y}$ as a function of x. 
%\as{questa \`e quella stazionaria l'abbiamo chiamata $u_{0y}(x)$ anche nella figura il label va modificato. 
%Stavo pensando anche che forse si potrebbero indicare le posizioni dei due shears anche nel plot con linee verticali per visualizzarli meglio, che dite? }
}
\label{fig:shearV}
\end{figure}

In our case, we used the following form for the function $U(x)$:
\begin{equation}\label{U}
U(x) = U_0 \left[  \tanh \left( \frac{x-L/4}{\Delta}\right) - \tanh \left( \frac{x-3L/4}{\Delta}\right) -1 \right]
\end{equation}
which gives a profile for $u_{0y}(x)$ with two shears localized at $x=L/4$ and $x=3L/4$. The bulk velocity is $u_{0y}\simeq U_0$ in the center $x=L/2$ of the spatial domain, while it is $u_{0y}\simeq -U_0$ on the two sides $x=0$ and $x=L$. In particular, we used the following values for the above parameters: $U_0=0.8$, $\Delta=2.5$. The corresponding profile for the bulk velocity $u_{0y}(x)$ is shown in Figure \ref{fig:shearV}. The value of $U_0$ has been chosen relatively large to reduce the {\it phase-mixing} time, and, consequently, the computation time. On the other hand, a large jump $\Delta u_y$ in the bulk velocity across the shear layers can lead to the development of the Kelvin-Helmholtz (KH) instability which would superpose to the wave dynamics. In order to prevent the KH instability to develop in our simulations, we have chosen $U_0$ sufficiently low to fulfil the condition $\Delta u_y < 2 c_A$. This choice stabilizes the system against the KH instability, at least in the MHD case. In particular, we have chosen $\Delta u_y \simeq 1.6 c_A$ ($c_A=1$ in code units). 
%{\op Se necessario, la parte su KH puo' essere ridotta drasticamente}

We notice that in the two shear regions the DF (\ref{f0}) departs from a shifted Maxwellian. In particular, the proton temperature is anisotropic: $T_{||} \ge T_\perp$, where $T_{||}$ and $T_\perp$ are the parallel and perpendicular temperature with respect to the magnetic field direction, respectively, and agirotropy features are also present \citep{malara18}. 
%{\fp{[Qual \`e il valore tipico di queste anisotropie/agirotropie?] }} \op{[Dipende da $\Delta$. Nel caso $\Delta=2.5 d_p$, sono dell'ordine 0.3 - 0.5 (vedi Malara 2018, Fig. 3)]}
Far from the velocity shears the proton temperature becomes isotropic and reaches the value $T_{0p}$, corresponding to the asymptotic thermal speed $v_{th0,p}$.
%\fm{Che ne pensate? Vi piace cos\`i?}. 

\subsection{Alfv\'enic perturbation}
At the initial time $t=0$ a perturbation is superposed on the above stationary configuration, with properties similar to a linearly-polarized large-scale Alfv\'en wave. Within MHD theory, an Alfv\'en wave is polarized in the direction perpendicular both to the background magnetic field ${\bf B}_0$ and to the wavevector ${\bf k}$. In our case, ${\bf B}_0$ is in the $y$ direction, while ${\bf k}$ varies in time due to {\it phase-mixing}, but remaining in the $xy$ plane. Therefore, we considered a perturbation polarized along the $z$ direction.
We considered a proton DF in the form: $f=f_0+f_1$, where the perturbed DF has the form of a Maxwellian shifted in the $v_z$ direction:
\begin{equation}\label{f1}
f_1(y,v_x,v_y,v_z)=\frac{n_1}{(2\pi )^{3/2} v_{th0,p}^3} 
\exp \left\{ -\frac{v_x^2 + v_y^2 + \left[ v_z - U_z(y)\right]^2}{2 v_{th0,p}^2} \right\}
\end{equation}
where $U_z(y)=\cos (k_0 y)$, with $k_0=2\pi /L$ corresponding to a wavelength $\lambda_y$ equal to the domain size $L$, and $n_1$ constant. We notice that $\lambda_y \gg d_p=1$. Therefore, the initial perturbation can be considered to be in an MHD regime. The total density is uniform and is given by 
\begin{equation}\label{ntot}
n = \int f_0(x,{\bf v}) d^3 {\bf v} + \int f_1(y,{\bf v}) d^3 {\bf v} = n_0 + n_1 = {\rm const.}
\end{equation}
This is coherent with the fact that an Alfv\'en wave does not involve density perturbations.
The bulk velocity is given by a weighted average between the stationary state bulk velocity ${\bf u}_0$ and the bulk velocity associated with the perturbation:
\begin{equation}\label{utot}
{\bf u}(x,y) = \frac{1}{n} \int {\bf v} f({\bf x},{\bf v}) d^3 {\bf v} = 
\frac{ n_0 u_{0y}(x) {\bf e}_y + n_1 U_z(y) {\bf e}_z}{n_0 + n_1}
\end{equation}
We have fixed the value $n_0=1$, while the parameter $n_1$ has been used to determine the amplitude $A_1=n_1/(n_0+n_1)$ of the initial Alfv\'enic perturbation.

\begin{table}
  \begin{center}
  \begin{tabular}{lcccccccccc}
%  \hline \hline
   Run  & $L$  & $n_0$  & $v_{th,p}$ & $\beta_p$ & $\Delta$ & $U_0$ & $k_0$ & $n_1$ & $A_1$ & $t_{max}$ \\[3pt] %\hline
 1  &    $16\pi$    &    $1$    &    $1$    &    $2$    &    $2.5$    &    $0.8$    &    $2\pi/L$     &    $0.01$    &    $0.0099$    &    80.0    \\[1pt]
 2 & $16\pi$ & $1$ & $1$ & $2$ & $2.5$ & $0.8$ & $2\pi/L$  & $0.1$ & $0.091$ & 100.0 \\ 
%\hline

  \end{tabular}
  \caption{Values of parameters used in the runs
%\tm{proporrei una tabella fatta così ma toglierei le due linee più esterne (cosa che non so fare) } \fp{per eliminare le linee basta commentare il comando 'hline'}
}
  \label{tab:params}
  \end{center}
\end{table}

In the MHD regime, velocity ${\bf u}_1$ and magnetic field ${\bf B}_1$ fluctuations are related by the expression ${\bf B}_1 = \mp (B_0/c_{A}) {\bf u}_1$, where $c_{A}=1$ is the normalized Alfv\'en velocity associated with the equilibrium structure and the upper (lower) sign corresponds to waves propagating in the direction of ${\bf B}_0$ (opposite to ${\bf B}_0$). Therefore, we have chosen the magnetic field initial perturbation as:
\begin{equation}\label{B1}
{\bf B}_1(y) = -\frac{n_1}{n_0 + n_1} U_z(y) {\bf e}_z = -A_1 U_z(y) {\bf e}_z
\end{equation}
%{\fp{[In questa formula forse conviene mettere direttamente $A_1$. Non dovrebbe apparire anche una radice della densit\`a da qualche parte (che forse \`e 1, perci\'o ininfluente)?]}}
%\op{\sout{being $v_{A0}=1$ in normalized units}}. 
Finally, the initial electric field ${\bf E}$ is determined by the generalized Ohm's law (\ref{ohm}).

\section{Numerical results}
We have performed numerical simulations using the HVM numerical code with the initial conditions specified in the previous section. Two simulations have been run, with two different amplitudes of the initial Alfv\'en wave: a low amplitude case $A_1=9.9\times 10^{-3}$ (corresponding to $n_1=0.01$, Run 1) and a moderate amplitude case $A_1=9.1 \times 10^{-2}$ (corresponding to $n_1=0.1$, Run 2). In both runs the electron temperature is equal to the proton temperature of the stationary state far from the shear layers: $T_e=T_{0p}$.
%{\fp{[Prima abbiamo detto che la temperatura degli ioni \'e anisotropa, quale temperatura si intende qui?]}}\fv{sì, si crea confusione, oltretutto credo che sia uguali all'inizio, dopo la temperatura protonica fa quello che vuole mentre quella elettronica resta fissa, giusto?} 
The proton $\beta_p$ is $\beta_p=2 v_{th,p}^2/c_A^2=2$ (i.e. $v_{th,p}=c_A=1$ in code units) and the typical thermal speed is the one evaluated far from the shear regions $v_{th,p}=v_{th0,p}$. We used a grid in the physical space of $N_x\times N_y$ points, with $N_x=512$ and $N_y=128$, while the grid in the velocity space has $71^3$ points. The maximum simulation time is indicated by $t_{max}$ for both runs. Other parameters have been specified in the previous section and are listed in Table \ref{tab:params}.

%\begin{table}
 % \begin{center}
%\def~{\hphantom{0}}
 % \begin{tabular}{lcccccccccc}
  % run\#  & $L$  & $n_0$  & $v_{th,p}$ & $\beta_p$ & $\Delta$ & $U_0$ & $k_0$ & $n_1$ & $A_1$ & $t_{max}$ \\[3pt]
 %Run 1  &    $16\pi$    &    $1$    &    $1$    &    $2$    &    $2.5$    &    $0.8$    &    $2\pi/L$     &    $0.01$    &    $0.0099$    &    80.0    \\
 %Run 2 & $16\pi$ & $1$ & $1$ & $2$ & $2.5$ & $0.8$ & $2\pi/L$  & $0.1$ & $0.091$ & 100.0

  %\end{tabular}
  %\caption{Values of parameters used in the runs}
  %\label{tab:params}
 % \end{center}
%\end{table}

\subsection{Run 1: small-amplitude perturbation}
We start by describing the results obtained in the small amplitude run (Run 1, $A_1 \simeq 0.01$). Due to the low value of the wave amplitude, nonlinear effects are negligible and the dynamics is completely dominated by the interaction of the wave with the inhomogeneity determined by the background velocity shear. Therefore, the purpose of this run is to clearly single out inhomogeneity linear effects, such as {\it phase-mixing}. In the following, the fluctuating part of any quantity $F({\bf x},t)$ is defined as $\delta F=F-\langle F \rangle_{D_{\bf x}}$, where angular parentheses indicate an average over the spatial domain $D_{\bf x}$.

%OP\centerline{\includegraphics[height=9cm,width=13cm]{bz_by.eps}}
\begin{figure}
  \centerline{\includegraphics[width=\textwidth]{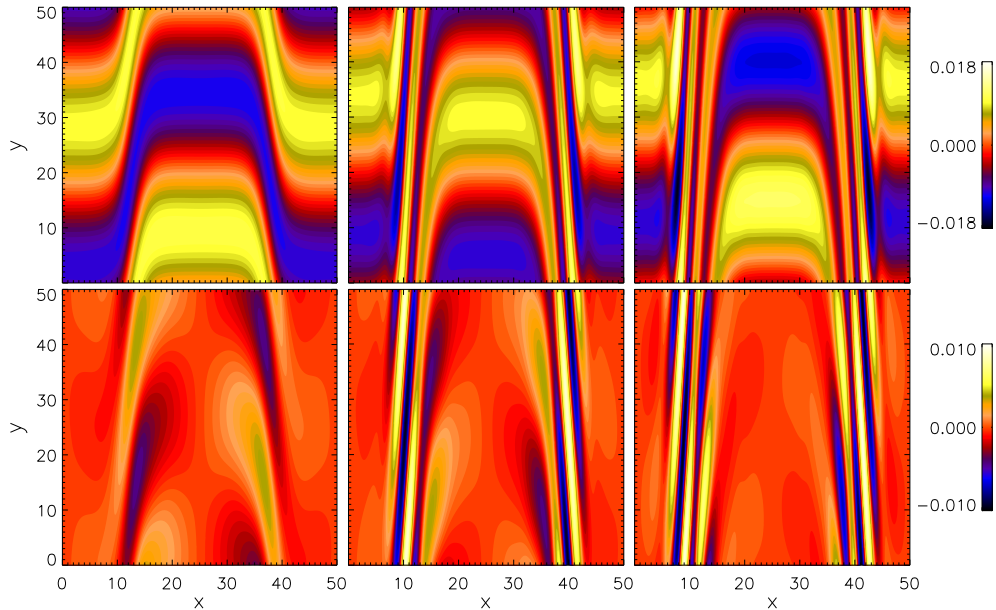}}
  \caption{2D plots in the $xy$ plane of fluctuating magnetic field components $\delta B_z$ (upper panels) and $\delta B_y$ (lower panels) at three different times: $t=20$ (left); $t=60$ (center); and $t=80$ (right). All plots refer to the low-amplitude run (Run 1).}
\label{fig:BzBy_small}
\end{figure}

In Figure \ref{fig:BzBy_small} the fluctuating magnetic field component $\delta B_z$ is plotted for three different times during the simulation (upper panels). At the initial time (not shown), only the $\delta B_z$ component, corresponding to the initial Alfv\'enic perturbation, is nonvanishing. The effects of {\it phase-mixing} on the time evolution of $\delta B_z$, due to space variations of the bulk velocity $u_y$, are clearly visible. The wave propagation velocity ${\bf v}_W$ in the simulation reference frame is the sum of the Alfv\'en velocity plus the bulk velocity ${\bf v}_W(x)=\left[c_A+u_y(x)\right] {\bf e}_y$. Accordingly, $v_W$ is larger in the center of the spatial domain than on the two sides. As a consequence, initially plane wavefronts are progressively bent in the two velocity shear regions, where the wavevector perpendicular component $k_x$ locally increases, while the wavelength decreases. Dispersive effects become effective for wavevectors of the order of $k_{disp}\simeq d_p^{-1} = 1$, corresponding to a wavelength $\lambda_{disp} \simeq 2\pi$. At later times in the simulation ($t \gtrsim 60$), the wavelength of the perturbation in the shear regions has decreased to $\lambda \sim \lambda_{disp}$. After that time, dispersive effects becomes relevant, at least within the velocity shear regions. In these regions, we observe that the magnetic fluctuation parallel component $\delta B_y$ (shown in Figure \ref{fig:BzBy_small}, lower panels) increases in time, until reaching values of the same order of $\delta B_z$ for times $t \ge 60$. In the MHD case, where dispersive effects are neglected, the same {\it phase-mixing} problem would give a null $\delta B_y$ for all times, implying that the generation of $\delta B_y$ is only caused by non-MHD dispersive effects. In fact, $\delta B_y$ grows only in the two shear regions, where the wavelength is small enough to make dispersive effects relevant.
In summary, in the velocity shear regions perturbations change their properties: the initial linear Alfv\'enic polarization is modified by the growth of a parallel component $\delta B_y$, the wavevector becomes nearly perpendicular ($k_{||} \ll k_\perp$) and of the order of the inverse proton inertial length. These evidences suggest that the initial Alfv\'enic perturbation has been locally converted into a KAW.

\begin{figure}
%\plottwo{figure3a.eps}{figure3b.eps}
  \centerline{\includegraphics[width=1.0\textwidth]{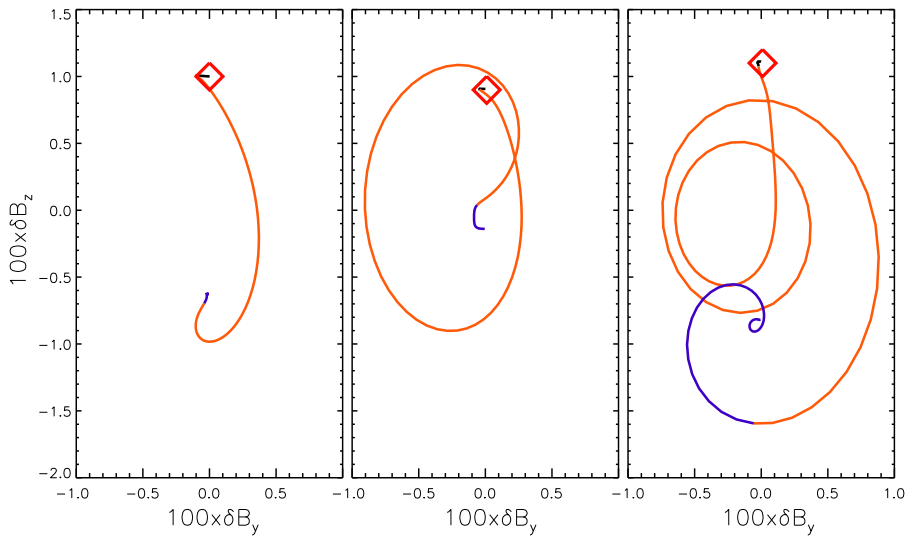}}
  \caption{Hodograms in the $\delta B_y$-$\delta B_z$ plane, relative to Run 1, at time $t=t_1=20$ along the line $L/2 \le x \le L$, $y=y_1=10$ (left panel); at time $t=t_2=40$ along the line $L/2 \le x \le L$, $y=y_2=45$(center panel); and at time $t_3=80$ along the line $L/2 \le x \le L$, $y=y_3=15$ (right panel). Black, orange and blue sections are relative to the intervals $L/2 \le x < 31$, $31 \le x < 44.5$, and $44.5 \le x \le L$, respectively. The red diamond indicate the point at $x=L/2$.}
\label{fig:hodo}
\end{figure}

In order to have a clear identification of the perturbations generated in the velocity shear regions, we have first considered the group velocity ${\bf v}_g$ of the waves. \citet{vasconez15} have shown that in the dispersive MHD all wave modes have a nonvanishing component of $v_{g\perp}$ in the direction perpendicular to ${\bf B}_0$. In particular, for fast magnetosonic waves it is $v_{g\perp} \gtrsim c_A$, while for slow magnetosonic and Alfv\'en waves it is $v_{g\perp} \ll c_A$ \citep{vasconez15}. In our simulation, we noticed that fluctuations generated in the shear regions propagate also perpendicularly to ${\bf B}_0$, in the direction from the center to the lateral parts of the spatial domain. However, we verified that the transverse propagation speed is one order of magnitude lower than the Alfv\'en speed. 
%To see this, we can compare the center and right panels of Figure \ref{fig:BzBy_small}: in a time interval $\Delta t=20$ the lateral displacement of the perturbation in the two shear regions is $\Delta x \sim 1$-$2$, while a transverse propagation speed of the order of $c_A=1$ would have produced a displacement $\Delta x \sim 20$. 
We conclude that such fluctuations cannot belong to the fast magnetosonic branch. 

To further discriminate among slow and Alfv\'enic fluctuations, we have examined how the orientation of $\delta {\bf B}$ varies along segments parallel to the $x$ axis. In the velocity shear regions these segments are quasi-parallel to the wavevector direction. An example is given in Figure \ref{fig:hodo}, where three hodograms are plotted: $\delta B_z$ vs $\delta B_y$. These are calculated at three different times: the left panel refers to the segment defined by $L/2 \le x \le L$, $y=y_1=10$ at time $t_1=20$, the center panel to the segment $L/2 \le x \le L$, $y=y_2=45$ at time $t_1=40$, while the right panel to the segment $L/2 \le x \le L$, $y=y_3=15$ at time $t_3=80$. The values $y_1$, $y_2$ and $y_3$ have been chosen such that $\delta B_z(x=L/2,y_{1,2,3},t_{1,2,3}) \simeq \max \left\{ \delta B_z(x=L/2,y,t_{1,2,3})\, , \, 0\le y \le L \right\}$. 
Different sections of the segments are indicated by different colors: black corresponds to the central homogeneous region ($L/2 \le x < 31$); orange corresponds to the shear region ($31 \le x < 44.5$); blue corresponds to the lateral homogeneous region ($44.5 \le x \le L$). The diamond indicates the values of $\delta B_y$ and $\delta B_z$ at the position $x=L/2$. 

Figure \ref{fig:hodo} can be used to determine the polarization of the magnetic perturbation in the $\delta B_y$-$\delta B_z$ plane, which is nearly perpendicular to the wavevector in the shear region. 
The left panel of Figure \ref{fig:hodo} refers to an early stage of time evolution. At that time, in the shear region (orange section of the curve) the wave polarization is still essentially linear, $z$-aligned, as the initial Alfv\'en wave. In fact, since the perturbation wavelength in the shear region is still larger than $d_p$ (see Figure \ref{fig:BzBy_small}, left panels), dispersive effects are not large enough to sensibly modify the initial polarization. In the same hodogram we also notice very small variations of the perturbation in the two homogeneous regions (black and blue sections of the curve); this is due to the fact that the hodogram is drawn along a segment that, in the homogeneous regions, is quasi parallel to wavefronts.

This latter feature is present also at the time $t=40$ (central panel of Figure \ref{fig:hodo}), but now the polarization in the shear region (orange section) has turned into clockwise elliptical. This polarization change is due to dispersive effects: they are now more relevant in consequence of the wavelength decrease induced by {\it phase-mixing}. The same polarization is also found in the shear region at $t=80$ (right panel); the orange section of the curve has a larges number of turns with respect to $t=40$ because {\it phase-mixing} has further decreased the perturbation wavelength. %{\fp{(Siamo sicuri che sia dovuto al fatto che la lunghezza d'onda sia ancora diminuita? A me sembra che la lunghezza d'onda sia rimasta costante e che si siano formati nel tempo più fronti d'onda. Però non sono convintissimo.)}}
Hodograms calculated along other parallel segments, which are not shown here, display the same behaviour. Considering wave modes within the dispersive MHD \citep{vasconez15},
%{\fp{[MHD o HMHD?]}}\fv{dispersive MHD non \'e HMHD?}, 
Alfv\'en and fast magnetosonic waves are both clockwise-elliptically polarized, while slow magnetosonic waves are counter-clockwise polarized. We can therefore exclude that fluctuations generated in the shear regions during system evolution belong to the slow magnetosonic branch. 

On the base of the above considerations about polarization and group velocity, the perturbations generated in the velocity shear regions belong to the Alfv\'en branch, i.e., they are KAWs.
Therefore, the {\it phase-mixing} mechanism acting in the shear regions gradually triggers a mode conversion from the initial Alfv\'en  wave to KAW fluctuations. Finally, we notice that a small amplitude clockwise-elliptically polarized fluctuation is also present in the lateral homogeneous region at time $t=80$ (Figure \ref{fig:hodo}, right panel, blue section); due to the nonvanishing $v_{g\perp}$, KAWs do not remain confined into the shear regions but tend to move to the side homogeneous regions. 

\subsection{Run 2: moderate-amplitude perturbation}
When increasing the initial perturbation amplitude, the effects of fluctuating fields on the proton DF become more relevant. In order to study such kinetic effects, being at the same time able to identify the characteristics of the propagation of KAWs during the system evolution, in Run 2 we have chosen a moderately larger amplitude of the initial perturbation $n_1 = 0.1$, corresponding to $A_1=0.091$. A larger amplitude would produce significant deformations of the initial equilibrium configuration,   making hard to clearly identify the nature of the transverse fluctuations produced by {\it phase-mixing}.

The time evolution of magnetic and velocity fields in Run 2 is qualitatively similar to that observed in Run 1 (see Figure \ref{fig:BzBy_small}). In particular, the initial Alfv\'en wave undergoes {\it phase-mixing} in the two velocity shear regions, locally generating small-scale perturbations with a quasi-perpendicular wavevector. Such perturbations are elliptically polarized in the sense of the Alfv\'en dispersive branch and slowly expand in the transverse direction, moving outside of the shear regions. Therefore, also in this moderate-amplitude case we can conclude that the initial Alfv\'en wave is converted into a KAW-like fluctuation by the interaction with the background inhomogeneity due to the velocity shear. 

The main differences between the small and moderate-amplitude cases consist in the modifications of the proton distribution function induced by fluctuations, which are more relevant for Run 2. In order to give a quantitative measure for such a phenomenon, we define the quantity $\eta$:
\begin{equation}\label{eta}
\eta ({\bf x},t) = \frac{1}{n({\bf x},t)} \sqrt{ \int_{D_{\bf v}} \left[ f({\bf x},{\bf v},t) - 
f_0({\bf x},{\bf v}) \right]^2 d^3{\bf v}}
\end{equation}
%\fv{Se ho capito bene dalla nostra ultima chiacchierata su skype bisogna eliminare la dipendenza temporale dalla funzione $f_0$ nella equazione (3.1). GIUSTO?}
%\op{[Ma questa quantita' e' esattamente $\epsilon$? Ossia, $f_0$ e' la Maxwelliana associata a $f({\bf x},{\bf v},t)$ o la VDF di equilibrio (Eq. (2.5)? Nel primo caso, lo chiamerei $\epsilon$. Nel secondo caso, visto che la VDF di equilibrio non e' Maxwelliana, non capisco a pieno la connessione di questo parametro con gli effetti cinetici (e potremmo anche sottostimarli perche' tutto potrebbe essere mescolato durante l'evoluzione). Inoltre, nel secondo caso, come entra la dipendenza temporale? Tramite densita', velocita' termica ma non velocita' di bulk, visto che in Eq. (2.5) $U$ e' una funzione data, giusto?]}
which represents the $L^2$ norm of the departure (in velocity space) between the proton DF $f$, at a position ${\bf x}$ and time $t$, and the unperturbed DF $f_0$ at the same position, normalized to the density. Since $\eta$ is a positive-definite quantity, it is useful to consider also its maximum value calculated over the spatial domain, at a given time $t$: $\eta_{max}(t)=\max_{D_{\bf x}} \left\{ \eta ({\bf x},t) \right\}$. The corresponding position is indicated by ${\bf x}_M(t)$: $\eta({\bf x}_M(t),t) = \eta_{max}(t)$. 
%At the initial time, the distribuition function $f$ has a well-defined expression; therefore, the initial value of $\eta_{max}$ can be analytically calculated, and it turns out to be $\eta_{max}(t=0)=\left[ n_1/(n_0 + n_1)\right] (2\sqrt{\pi} v_{th,p})^{-3/2} \simeq 8\times 10^{-3}$. The calculation of $\eta_{max}(t=0)$ is reported in the Appendix. 

\begin{figure}
  \centerline{\includegraphics[width=0.6\textwidth]{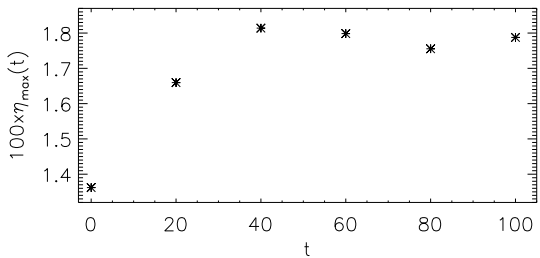}}
  \caption{The quantity $\eta_{max}$ is plotted as a function of time $t$, for the moderate-amplitude run (Run 2). 
%\fp{[Forse qui sarebbe utile graficare anche i valori di $\eta$ per RUN1, o quanto meno citare nel testo il valore tipico]}
}
\label{fig:etamax}
\end{figure}

In Figure \ref{fig:etamax} the quantity $\eta_{max}(t)$ is plotted as a function of time, for Run 2. At the initial time it is $\eta\ne 0$, due to the DF initial perturbation $f_1$ which is superposed on the stationary DF $f_0$. We can see that the departure of the DF $f$ from  $f_0$ slightly increases in time, indicating a progressive modification of the DF due to the effects of the perturbation. The growth of $\eta_{max}(t)$ saturates at the time $t\simeq 40$, remaining roughly constant after that time. The saturation value is $\eta_{max,sat} \simeq 1.8\times 10^{-2}$, which is $\simeq 1.3\, \eta_{max}(t=0)$. The relatively low increase of $\eta_{max}$ is partially due to the moderate amplitude of the perturbation and partially to the large value of $\beta_p$. In fact, larger values of $\beta_p$ corresponds to larger thermal energy in the proton population, when compared to the energy associated with the perturbation. We expect to find an increase of $\eta_{max}$ more relevant than in the present case, when larger perturbation amplitudes and/or lower $\beta_p$ are considered. 
%\op{[La crescita e' davvero minima, speriamo il referee non si ci appigli. Probabilmente va sottolineato che non si parte da zero visto che inizialmente la VDF non e' Maxwelliana]}
%\fv{in Figura 4, bisogna togliere tutto il bianco e suggerisco di sovrapporre l'evoluzione di eta anche nel caso con ampiezza 0.01, per fare vedere che in questo caso rimane circa costante.}

\begin{figure}
  \centerline{\includegraphics[width=\textwidth]{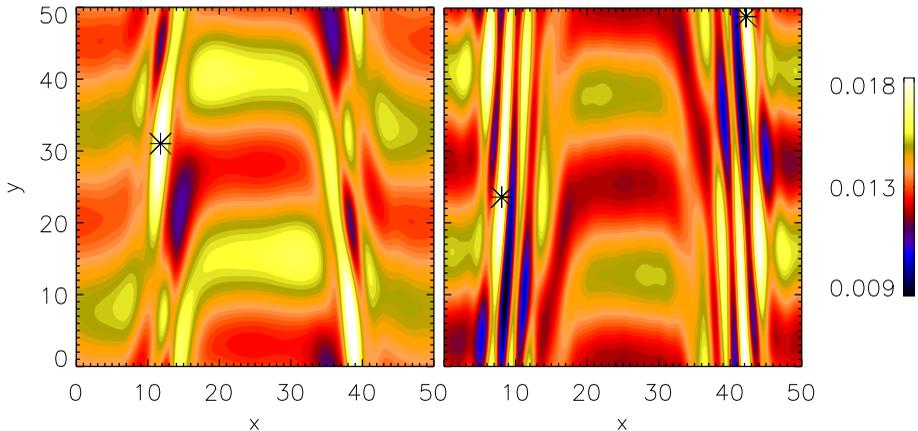}}
  
  \caption{2D plots of the quantity $\eta({\bf x},t)$ in the $xy$ plane, at time $t=40$ (left panel) and $t=100$ (right panel), for the moderate-amplitude run (Run 2). Asterisks indicate the positions where $\eta$ is maximum at the given time.}
\label{fig:eta}
\end{figure}

In Figure \ref{fig:eta}, 2D plots of $\eta({\bf x},t)$ in the $xy$ plane are shown at times $t=40$ and $t=100$. Asterisks indicate the location ${\bf x}_M$ where $\eta$ attains the maximum value $\eta_{max}$ for the given time. From this figure we see that the largest departures from the background DF $f_0$ are mainly localized within the velocity shear regions, where the smallest wavelengths form in the perturbation. The spatial structure of $\eta({\bf x},t)$ appears to be influenced by {\it phase-mixing}, indicating that the DF is also spatially modulated on a small transverse scale. Therefore, departures from the unperturbed proton DF are strictly related to the generation of small scales, of the order of $d_p$.
  
\begin{figure*}
\centering
\begin{minipage}{0.9\textwidth}
   \centering
\includegraphics[width=0.9\textwidth]{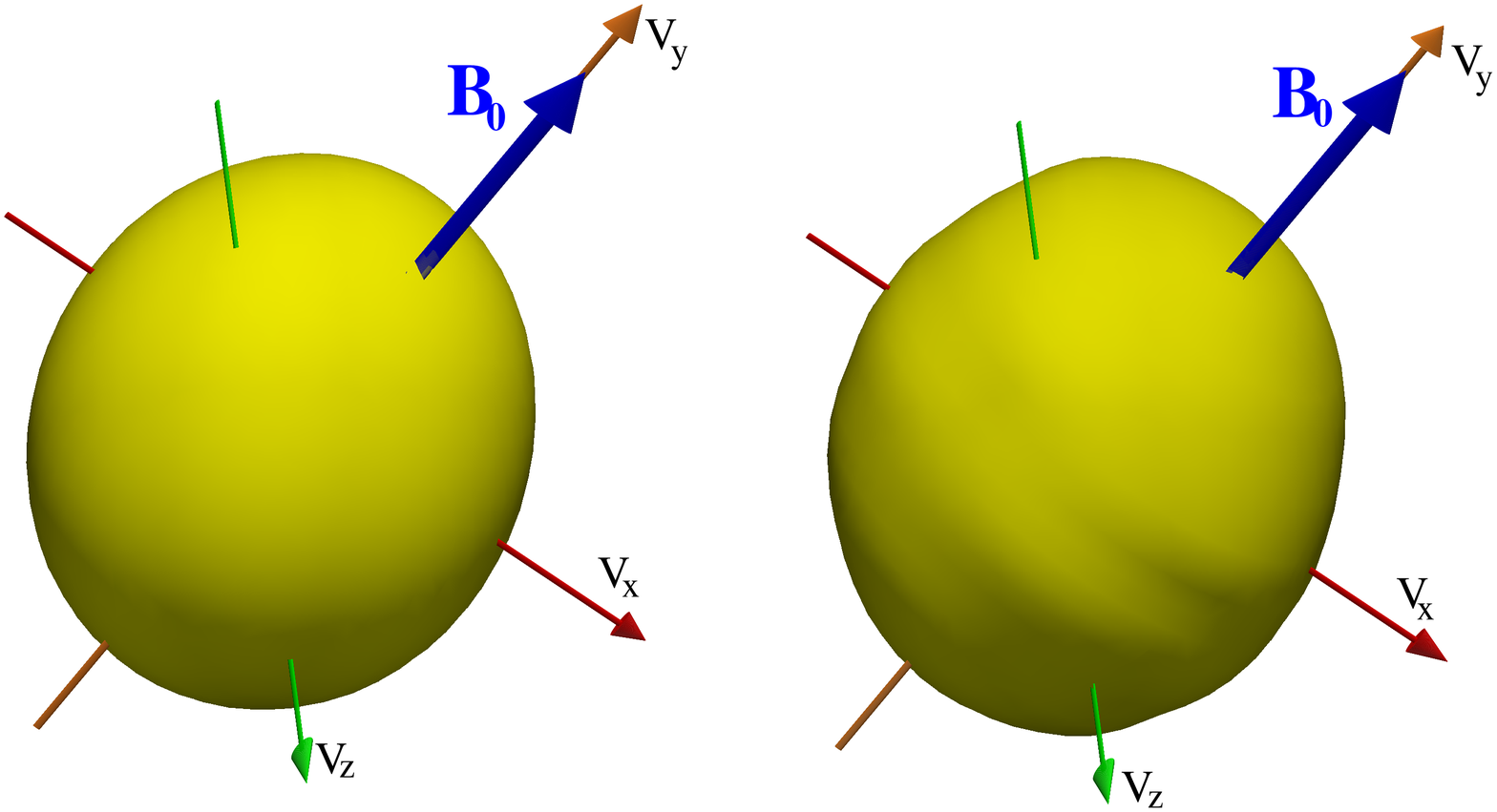}
  \end{minipage}
% \hfill
 \begin{minipage}{0.9\textwidth}
 \centering
   \includegraphics[width=0.9\textwidth]{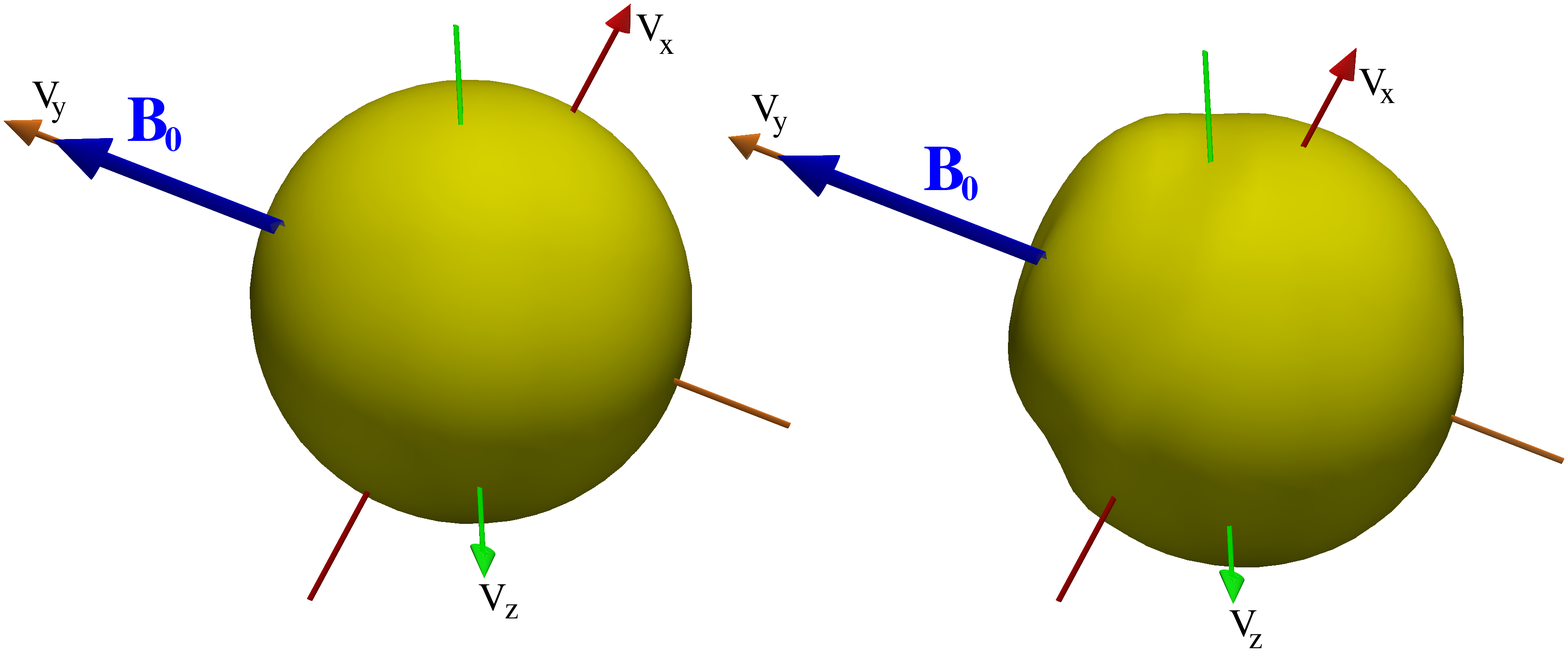}
 \end{minipage}
    \caption{Isosurfaces of the 3D proton velocity DF $f({\bf x}_M(t),{\bf v},t)$ (right panels), and of the stationary DF $f_0({\bf x}_M(t),{\bf v})$ (left panels), at time $t=40$ (upper panels) and $t=100$ (lower panels), calculated in the spatial position ${\bf x}_M(t)$ where $\eta=\eta_{max}$ at the given time. Isosurfaces are relative to the values $f=f_0=0.008$. Plots refer to the moderate-amplitude run (Run 2).}
\label{fig:DF3D}
\end{figure*}

\begin{figure}
  \centerline{\includegraphics[width=0.7\textwidth]{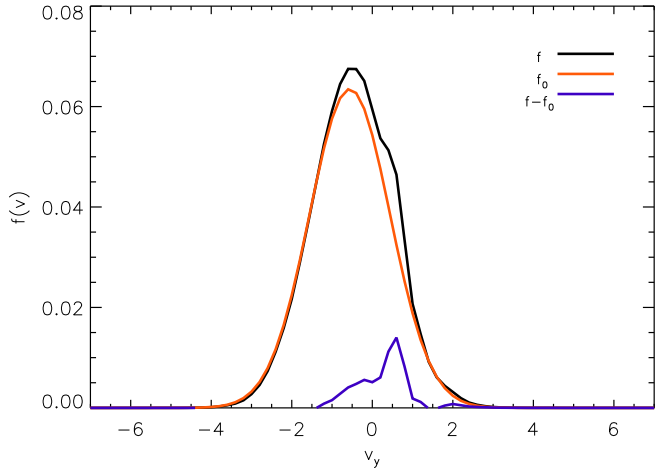}}
  \caption{Profiles of the proton distribution functions $f$ (black line), $f_0$ (orange line), and of the difference $f-f_0$ (blue line) are plotted as functions of $v_y$, for $v_x=v_z=0$, at the space position ${\bf x}_M(t)$ at time $t=100$. Plots refer to the moderate-amplitude run (Run 2).}
\label{fig:fcut}
\end{figure}

3D iso-surface plots of the proton DF $f({\bf x}_M(t),{\bf v},t)$ in velocity space are shown in Figure \ref{fig:DF3D} for $t=40$ and $t=100$, at the position ${\bf x}_M(t)$. For comparison, the unperturbed DF $f_0({\bf x}_M(t),{\bf v})$ is plotted at the same position (Figure \ref{fig:DF3D}, left panels). The DF modifications become evident by comparing $f$ and $f_0$. At time $t=40$, such variations appear as a modulation in form of rings on the considered isosurface (Figure \ref{fig:DF3D}, upper-right panel), which are co-axial with the direction $y$ of the background magnetic field ${\bf B}_0$, indicated by a blue arrow. We can also notice that the unperturbed DF $f_0$ departs from a Maxwellian, the parallel temperature $T_{p||}$ being larger than the perpendicular one $T_{p\perp}$ \citep{malara18}. At time $t=100$, the DF $f$ shows a peculiar feature where $\eta=\eta_{max}$; namely, a ``bulge" is present in the isosurface (Figure \ref{fig:DF3D}, lower-right panel), indicating the presence of a sub-population of protons moving along ${\bf B}_0$ faster than the core particles. The existence of this beam of accelerated protons can also be seen in Figure \ref{fig:fcut}, where a ``cut" of the DF shown in Figure \ref{fig:DF3D} for $t=100$ is plotted in velocity space as a function of $v_y$, for $v_x=v_z=0$ (black line). In the same figure, the corresponding profile of the unperturbed DF $f_0$ is plotted for comparison (orange line), along with the difference $f-f_0$ of the two profiles (blue line). The proton beam is evident also in Figure \ref{fig:fcut}; in particular, the difference in velocity between the beam and core population along the ${\bf B}_0$ direction is $\Delta v_y\simeq c_A=1$.

\begin{figure}
  \centerline{\includegraphics[width=0.6\textwidth]{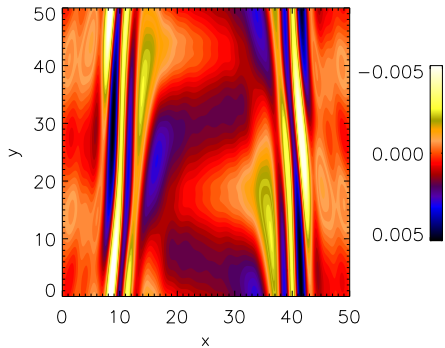}}
  \caption{2D plot of the electric field parallel component $E_y$ in the $xy$ plane, calculated at time $t=100$ for the moderate-amplitude run (Run 2).}
\label{fig:Ey}
\end{figure}

Concerning the origin of the beam, we observe that KAWs are characterized by a component of the electric field $E_y$ parallel to the background magnetic field ${\bf B}_0$ \citep[e.g.,][]{hollweg99}. Such a component is vanishing in the limit of ideal MHD and can be responsible for particle energization.
In Figure \ref{fig:Ey} a 2D plot of the parallel electric field component $E_y$ at the time $t=100$ is shown. Figure \ref{fig:Ey} shows that $E_y$ is localized in the region where the smallest spatial scales are present in the perturbation, while it is negligible outside. Moreover, the spatial pattern of $E_y$ is similar to that of other perturbed quantities, and the wavelength in the parallel $y$ direction is $\lambda_{||}=2\pi/k_0=L$. As a consequence, we can define a sort of electric potential $\phi$, such as $E_y=-\partial \phi/\partial y$. Of course, the dependence of $\phi$ on $y$ is also periodic, with a period equal to $L$. Protons, in their motion, feel this periodic potential and can interact with it. 
%\fp{[qui stiamo dicendo che siamo in un caso quasi elettrostatico in cui le variazioni di B sono piccole rispetto al campo guida?]}
In particular, part of the ion population can remain trapped into the potential well associated with the spatially-modulated $E_y$ component. For a given particle kinetic energy, this trapping is more probable for protons which move quasi-parallel to ${\bf B}_0$, i.e., with $v_x$ , $v_z \ll v_y$; this latter condition corresponds to the cut in Figure \ref{fig:fcut}. The potential well co-moves with the wave along the magnetic field with a velocity $\simeq c_A$ in the plasma bulk reference frame, which explains why trapped protons have an average velocity of $u_y(x)+c_A$. This process can account for the formation of the DF features observed in Figure \ref{fig:fcut}.

Indeed, in previous simulations of Alfv\'en wave {\it phase-mixing} generated by magnetic field inhomogeneities, the presence of ion beams moving along the magnetic field at the local Alfv\'en speed has been observed by \citet{vasconez15,valentini17}. These authors have shown that the origin of such beams is the parallel electric field associated with KAW-like fluctuations, generated in the inhomogeneity regions by the {\it phase-mixing} mechanism. It is likely that the same physical process is responsible for the creation of suprathermal ion beams also in the case considered in the present paper. 

\section{Conclusions}
In this paper we have studied the evolution of an Alfv\'en wave in a collisionless plasma, propagating in a stationary configuration characterized by a sheared flow with a uniform magnetic field parallel to the velocity field. The study has been carried out numerically, by means of the HVM code. We considered a width of the shear layers of the order of the proton inertial length $d_p$. In order to properly describe the unperturbed configuration, we used an exact analytical solution suitable for the HVM approach \citep{malara18}. The physical mechanism acting on the time evolution of the perturbation is {\it phase-mixing}, induced by transverse variations of the plasma bulk velocity, which bends wavefronts and generates increasingly small scales in the transverse direction. Though the initial perturbation is in a MHD regime, i.e., the wavelength $\lambda$ is initially much larger than proton scales, later in time $\lambda$ locally becomes of the order of $d_p$. Once this stage is reached, both dispersive and kinetic effects come into play and the initial Alfv\'en wave is converted into a KAW. The identification of the perturbation within the shear region as a KAW has been done considering both its polarization, which is turned from linear into clockwise-elliptical, and the transverse group velocity, which is nonvanishing but much less than the Alfv\'en speed. 
 
In the moderate amplitude run, properly kinetic effects have been observed, such as modifications of the initial stationary distribution function. In the considered case, variations of the DF are not particularly large, due both to the moderate perturbation amplitude and to the large value of the proton beta parameter ($\beta_p=2$); in fact, larger values of $\beta_p$ correspond to a lower ratio of the perturbation energy to the bulk thermal energy. We decided not to increase further the amplitude of the initial perturbation as for larger amplitudes the wave features of the KAW are no longer easily identified.

Nevertheless, an interesting feature has been observed, namely, the generation of a beam of suprathermal protons moving along the magnetic field with a speed given by the sum of the Alfv\'en velocity and the local bulk velocity. The existence of such a beam can be related to an electric field component $\delta E_{||}$ parallel to the magnetic field, which characterizes KAWs, and which has been indeed observed in our simulations in the velocity shears regions, where KAWs are generated. We notice that the presence of magnetic-field aligned proton beams, is often observed in the solar-wind plasma \citep{marsch06}. Therefore, our results suggest that such a feature could be related to the presence of KAW-like fluctuations in the solar wind.

Most of the behaviour and features observed in the present simulations have been also found in another situation in which an Alfv\'en wave propagate in a background configuration where the magnetic field ${\bf B}_0$ is inhomogeneous, with spatial variations perpendicular to ${\bf B}_0$, at scales comparable with the proton inertial length \citep{vasconez15,pucci16,valentini17}. In that case, {\it phase-mixing} acting on the Alfv\'en wave is due to spatial variations of the Alfv\'en velocity. The results found in the latter studies and these obtained in the present investigation can be considered to be complementary. We can conclude that the observed phenomenology is due to {\it phase-mixing} at proton scales, regardless of the origin of such a process (magnetic field and/or bulk velocity inhomogeneities); in other words, as long as the propagation velocity of Alfv\'en waves is spatially inhomogeneous in the direction perpendicular to the background field, these can be successfully converted into KAWs. 
%\fp{[qui suggerirei che l'ultima frase sia messa pi\`u in rilievo. Di fatto stiamo dicendo l'unica cosa che conta è una disomogeneit\`a nella velocit\`a di propagazione, non \`e necesserio n\'e uno shear di B, n\'e uno shear di v, per\`o è sufficiente o uno shear di B o uno shear di v.]} \fv{Sono d'accordo con Ciccio P.}.

As a final remark, we observe that the present results have been derived within a simplified configuration, where it is possible to distinguish a single well-defined wave from the inhomogeneous background. Nevertheless, our results give some indications on the possibility of generating KAWs in more complex contexts, like in a turbulent setting. In fact, the wave-inhomogeneity coupling considered in this investigation is of a similar nature as the nonlinear coupling between fluctuations which generate the turbulent cascade in MHD: in a turbulent regime fluctuations with two different $k$s couple to produce a third one at higher $k$, while in the configuration here considered a fluctuation with a well defined $k$ couples with the $k$ coming from the inhomogeneity. 

We therefore conclude that the phenomenon we studied is closely related to the mechanism that favors perpendicular spectral transfer in the nonlinear cascade \citep[e.g.,][]{shebalin83}. Moreover, our results give a positive indication about the possibility of generating KAW-like fluctuations at proton scales within a turbulent cascade, as suggested by solar wind observations \citep{bale05,sahraoui12}.

\section*{Acknowledgements}
The authors are grateful to Prof. W.H. Matthaeus for many useful discussions about the subject of the paper. The work by F. Pucci has been supported by Fonds Wetenschappelijk Onderzoek – Vlaanderen (FWO) through the postdoctoral fellowship 12X0319N. This paper has received funding from the European Union’s Horizon 2020 research and innovation program under Grant Agreement No. 776262 (AIDA, www.aida-space.eu)

\end{document}